\documentclass{emulateapj}
\usepackage{lscape}
\usepackage{color}

\shorttitle{autoMOOG: Abundance Fitting of VMP Stars}
\shortauthors{Marsteller et al.}

\begin{document}
\title{Automated Determination of [F\lowercase{e}/H] and [C/F\lowercase{e}] from
  Low-Resolution Spectroscopy}
\author{B. Marsteller\altaffilmark{1,2}, T. C. Beers\altaffilmark{1},
  T. Sivarani\altaffilmark{1,3}, S. Rossi\altaffilmark{4},
  V. Placco\altaffilmark{4}, G. R. Knapp\altaffilmark{5},
  J. A. Johnson\altaffilmark{6}, S. Lucatello\altaffilmark{7}}

\altaffiltext{1}{Department of Physics \& Astronomy, CSCE: Center for the Study of Cosmic Evolution, and JINA: Joint Institute for Nuclear Astrophysics,  Michigan State University, East Lansing, MI 48824, USA; bmarstel@uci.edu}
\altaffiltext{2} {Department of Physics \& Astronomy, University of California, Irvine, 2180 Frederick Reines Hall, Irvine, CA 92697-4565, USA}
\altaffiltext{3} {Department of Astronomy, University of Florida, 211 Bryant Space Science Center, Gainesville, FL 32611-2055, USA}
\altaffiltext{4}{Departamento de Astronomia, Instituto de Astronomia, Geof\'isica e Ci\^encias Atmosf\'ericas, Universidade de S\~ao Paulo, Rua do Mat\~ao 1226, 05508-900 S\~ao Paulo, Brazil}
\altaffiltext{5}{Department of Astrophysical Sciences, Princeton University, Princeton, NJ 08544}
\altaffiltext{6}{Department of Astronomy, Ohio State University, 140 W. 18th Avenue, Columbus, OH 43210}
\altaffiltext{7}{Osservatorio Astronomico di Padova, Vicolo dell'Osservatorio 5, 35122 Padua, Italy and Excellence Cluster Universe, Technische Universit\"at M\"unchen, Boltzmannstr. 2, D-85748, Garching, Germany}

\begin{abstract}
We develop an automated spectral synthesis technique for the
estimation of metallicities ([Fe/H]) and carbon abundances ([C/Fe])
for metal-poor stars, including Carbon-Enhanced Metal-Poor (CEMP)
stars, for which other methods may prove insufficient.  This
technique, autoMOOG, is designed to operate on relatively strong
features visible in even low- to medium-resolution spectra, yielding
results comparable to much more telescope-intensive high-resolution
studies.  We validate this method by comparison with 913 stars which
have existing high-resolution and low- to medium-resolution spectra,
and that cover a wide range of stellar parameters.  We find that at
low metallicities ([Fe/H] $\lesssim -2.0$), we successfully recover
both the metallicity and carbon abundance, where possible, with an
accuracy of $\sim0.20$ dex.  At higher metallicities, due to issues of
continuum placement in spectral normalization done prior to the
running of autoMOOG, a general underestimate of the overall
metallicity of a star is seen, although the carbon abundance is still
successfully recovered.  As a result, this method is only recommended
for use on samples of stars of known sufficiently low metallicity.
For these low-metallicity stars, however, autoMOOG performs much more
consistently and quickly than similar, existing techniques, which
should allow for analyses of large samples of metal-poor stars in the
near future.  Steps to improve and correct the continuum placement
difficulties are being pursued.
 \end{abstract}

 \keywords{Galaxy: halo -- stars: abundances -- stars: carbon --
   stars: Population III -- techniques: spectroscopic}

\section{INTRODUCTION}
One of the most intriguing results from modern surveys of metal-poor
stars, such as the HK \citep{B85,B92} and Hamburg/ESO surveys
\citep[HES;][]{Christlieb}, is the large fraction of Carbon-Enhanced
Metal-Poor (CEMP; [C/Fe] $\geq +1.0$) stars identified amongst the
lowest metallicity stars.  The exact fraction of CEMP stars at low
metallicity has been debated in the literature, and is variously
reported as $\sim$10\% \citep{Frebel}, $\sim$15\% \citep{Cohen} or
$\sim$20\% \citep{Luc} of stars with [Fe/H] $\leq -2.0$; $\sim$20\%-25\%
of stars with [Fe/H] $\leq -2.5$ \citep{Mars}; or nearly 40\% of stars
with [Fe/H] $\leq -3.5$ \citep{BaC}.  The fraction of CEMP stars is
expected, on theoretical grounds, to increase with declining
metallicity \citep{Tumlinson}.  The evolutionary states of sample
stars can also play a significant role, due to the mixing-induced
dilution of carbon \citep{A06,Luc}.

To better specify the changing CEMP fraction with metallicity, much
larger samples of stars are required.  Existing studies have generally
been limited to (small) samples of low-metallicity stars for which
high-resolution spectral abundances have been obtained \citep[with the
exception of][]{Frebel}.  Such spectra are time consuming to obtain
and simply fail to exist in large enough samples.  Lower-resolution
spectra are available in sufficient quantity, but the techniques used
to determine abundances are more inconsistent across parameter space,
and largely based on empirical calibrations
\citep[e.g.,][]{B99,Rossi}.

Here we discuss the development of an automatic spectral synthesis
fitting routine, autoMOOG, which enables rapid spectral fitting of the
low- to medium-resolution spectra available in existing and future
large samples of metal-poor stars.  Papers in preparation will use
this technique, and others, on a much larger sample of stars than has
been studied to date, to determine the frequency of carbon enhancement
as a function of metallicity.  autoMOOG will also be vital for
identifying the most useful and interesting targets from large data
sets for future high-resolution follow-up.

This paper is outlined as follows.  Section 2 presents a discussion of
the current techniques and a description of the new method that has
been developed.  Error analysis, performed on the program using
synthetic spectra, is presented in Section 3.  Section 4 considers
validation of the autoMOOG approach using medium-resolution spectra of
stars with available high-resolution determinations of [Fe/H] and
[C/Fe].  The important question of the specification of upper limits
on [C/Fe] for warmer stars (where the prominent C features are quite
weak) is addressed in Section 5.  Section 6 presents brief conclusions
and a discussion of future work anticipated for refinement of
autoMOOG.

\section{METHODOLOGY}
For metal-poor stars with only low- to medium-resolution spectroscopy
available, stellar parameters such as metallicity ([Fe/H]) and carbon
abundance ([C/Fe]) have primarily been estimated from broadband colors
and line indices measuring the strength of the Ca\textsc{ii} K line
and the CH $G$ band--the KP and GP indices, respectively
\citep{B99,Rossi}. However, for CEMP stars, the presence of strong
carbon features may affect the measured colors of a star, in
particular the $B$ magnitude. \citet{Rossi} have attempted to account
for this by using $J - K$ colors instead of $B - V$ used by
\citet{B99}. Strong carbon features may also affect the measurement of
the continuum in a side band of the KP index \citep[as first indicated
by][]{Cohen}. Such effects could lead to an underestimation of the
stellar metallicity.  Very strong carbon features can also
``overflow'' into the sidebands used for the measurement of the GP
index, confounding estimates of [C/Fe].

It is useful to explore different approaches that should be relatively
immune to these effects.  One such method involves the use of model
stellar atmospheres to generate synthetic spectra, so that detailed
fits of a given observed spectrum may be made, rather than suffer the
loss of information inherent in techniques based on line indices.  We
have explored using the spectrum analysis code MOOG \citep{Sneden}, a
grid of ATLAS9 1D, LTE stellar atmosphere models and atomic and
molecular line lists \citep[including lines of CH, CN, and
C$_2$;][]{CaK} to carry out this approach.  We have developed a
routine to automatically run MOOG and minimize the residuals of the
resulting model spectra when compared against certain critical regions
of the observed data.

Using the line index methods in concert with other techniques
\citep[see][]{Lee}, T$_{eff}$ and log $g$ values are adopted and
initial estimates are determined for [Fe/H] and [C/Fe].  From
our grid of models (with spacing of 100 K in T$_{eff}$, 0.2 in log
$g$, and 0.5 in [M/H], or overall solar-scaled metallicity), we select
the model that most closely matches our starting parameters.  This
model is used within MOOG, where individual abundances are then varied
to the nearest 0.05 dex to fit a full line profile.  If the
metallicity is varied significantly, the next appropriate model in the
grid is used instead.  Any element not currently being fit is given
the model solar-scaled value.  However, the user may specify other
individual elements to vary as well, generally by a fixed offset from
the scaled value.  Thus, for example, it would be possible to use CH
to determine [C/Fe], and then use that value to fit CN to determine
[N/Fe].

As in the line index method, the Ca \textsc{ii} K line at 3933 {\AA}
is used to solve for the metallicity of a star.  At low- to
medium-resolution, no iron lines are strong enough to be cleanly
measured, so calcium is used as a proxy for metallicity.  However,
unlike the line index method, which empirically calibrates the strength
of the KP index to the metallicity, here the Ca \textsc{ii} K line is
fit to obtain a calcium abundance.  Since calcium is an $\alpha$-element,
an appropriate, metallicity-dependent offset, to account for possible
$\alpha$-enhancements, must be used when determining the metallicity
of a star.  Following \cite{B99}, we use an $\alpha$-enhancement
applicable for Milky Way stars of:
\begin{equation}
[\alpha/\textrm{Fe}] = \left\{ \begin{array}{ll}
  0 & \textrm{if [Fe/H]} \geq 0\\
  -0.267*[\textrm{Fe/H}] & \textrm{if} -1.5 \leq [\textrm{Fe/H}] < 0\\
  0.4 & \textrm{if [Fe/H]} < -1.5\\
\end{array} \right. \label{eq:alpha}
\end{equation}
The initial guess for [C/Fe] is also included in our metallicity fit to
address the effect pointed out by \citet{Cohen}.

Using the new, more precisely determined metallicity, [C/Fe] can
similarly be fit using the CH $G$ band at 4304 \AA.  However, for warmer
stars, those with a lower carbon abundance, or those that are very
metal-poor, the $G$ band may be too weak to be detected.  In these
cases, an upper limit for [C/Fe] is determined instead.  In theory,
with an updated [C/Fe], it may be necessary to iterate on the fit to
the Ca \textsc{ii} K line.  However, extensive testing revealed that
the metallicity determined using the input [C/Fe] and that obtained
from the refined [C/Fe] varied by negligible amounts (if at all),
suggesting that such iteration is not required at this precision.

\begin{figure}
  \figurenum{1}
  \plotone{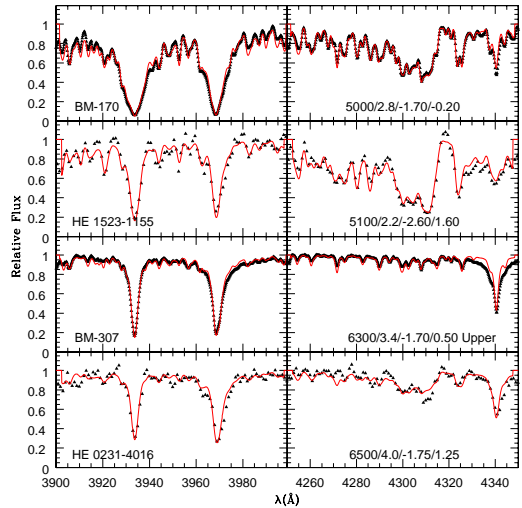}
  \caption{Example of the fits (line) produced by autoMOOG of the Ca
  \textsc{ii} K region (left) and of the $G$ band (right) compared to
  the data (triangles).  The numbers in the right panels correspond to
  the T$_{eff}$, log $g$, [Fe/H], and [C/Fe], respectively, of the
  star.  Note the higher resolution of the BM stars compared to the
  HERES sample, showing that a range of spectral resolutions can
  easily be fit.  These stars also show the effect of temperature and
  abundances on the appearance of the Ca \textsc{ii} K line and the $G$
  band, as well as the appearance of an upper limit determination.
    \label{fig:fits}}
\end{figure}

Once this approach was worked out by fitting the abundances of a number
of stars by hand, a program, autoMOOG, was developed to iteratively
vary abundances and automatically minimize the $\chi^2$ of the
synthetic spectra produced by MOOG.  autoMOOG allows the user to
select any wavelength region for the comparison.  For the results
presented here, a fit was determined using the ranges 3910$-$3960 {\AA} for
the Ca \textsc{ii} K line, and 4275$-$4330 {\AA} for the CH $G$ band.  An
example of the fits produced by autoMOOG can be seen in Figure
\ref{fig:fits}.

As part of the automation, autoMOOG also uses the region around the Ca
\textsc{ii} K line to estimate the observed radial velocity, which is
used to align the data with the model spectra.  The velocity precision
is directly proportional to the spectral resolution, which for our
test spectra ($R \sim 4000$) is typically $\lesssim$10 km s$^{-1}$.
Continuum subtraction and spectral renormalization occur prior to the
running of autoMOOG.  Assuming this renormalization to be valid,
autoMOOG applies a small additive offset to the local continuum level
of the normalized, continuum-subtracted observed spectra such that it
matches the local continuum level of the comparison synthetic spectra
without altering the effective line strength.  As we will see in
Section 4, this assumption likely becomes less valid at higher
metallicities, where the apparent continuum is significantly
suppressed in the vicinity of the Ca \textsc{ii} line, leading to
incorrect continuum subtraction.

\section{ERROR ANALYSIS}
To estimate the errors on the derived abundances, a set of noiseless
synthetic spectra across the grid of parameters was generated.  Since
these spectra had known parameters, the abundance determinations could
be directly compared with the proper values.  These spectra were first
run through autoMOOG to determine any intrinsic scatter in the
estimates.  These errors, and their rough dependence on T$_{eff}$ and
log $g$, are listed in Table \ref{tbl:error}.  The errors are roughly
constant across both of these parameters, with the exception of very
cool giants where the scatter increases.  Also, note that the minimum
errors are limited by the resolution of the abundance grid (0.05 dex).

\addtolength{\tabcolsep}{-4pt}
\begin{deluxetable}{cccccccccc}
\tablecaption{Abundance Errors, and Their Dependence on T$_{eff}$ and log $g$\label{tbl:error}}
\tablewidth{0pt}
\tablehead{
\colhead{} & \multicolumn{4}{c}{[Fe/H]} && \multicolumn{4}{c}{[C/Fe]}\\\cline{2-5} \cline{7-10}
\colhead{Bin} & \colhead{$\sigma_{\mathrm{Int}}$} & \colhead{$\sigma_{\mathrm{T}_{eff}}$\tablenotemark{1}} & \colhead{$\sigma_{\mathrm{log }g}$\tablenotemark{2}} & \colhead{$\sigma_{\mathrm{Total}}$\tablenotemark{3}} &  & \colhead{$\sigma_{\mathrm{Int}}$} & \colhead{$\sigma_{\mathrm{T}_{eff}}$\tablenotemark{1}} & \colhead{$\sigma_{\mathrm{log} g}$\tablenotemark{2}} & \colhead{$\sigma_{\mathrm{Total}}$\tablenotemark{3}}}
\startdata
 \multicolumn{10}{c}{Errors in Bins of T$_{eff}$}\\
 \hline
 4000 K & 0.162 & 0.306 & 0.236 & 0.351 && 0.267 & 0.238 & 0.207 & 0.267\\
 5000 K & 0.058 & 0.168 & 0.073 & 0.174 && 0.058 & 0.107 & 0.064 & 0.111\\
 6000 K & 0.093 & 0.139 & 0.070 & 0.139 && 0.088 & 0.097 & 0.065 & 0.097\\
 7000 K & 0.041 & 0.123 & 0.046 & 0.125 && 0.033 & 0.104 & 0.085 & 0.130\\
 \hline
 \multicolumn{10}{c}{Errors in Bins of log $g$}\\
 \hline
 4.0 & 0.124 & 0.172 & 0.112 & 0.172 && 0.153 & 0.153 & 0.119 & 0.153\\
 4.6 & 0.139 & 0.154 & 0.139 & 0.154 && 0.152 & 0.161 & 0.141 & 0.161\\
 5.0 & 0.180 & 0.254 & 0.172 & 0.254 && 0.174 & 0.168 & 0.143 & 0.174\\
\enddata
\tablenotetext{1}{This error is the combined errors due to the
intrinsic error and due to the uncertainty in the input effective
temperature.  Since the individual errors are non-Gaussian, there is
no simple way to disentangle them.}
\tablenotetext{2}{This error is due to the combination of the
intrinsic error and the uncertainty in the input surface
gravity.}
\tablenotetext{3}{This is an estimate of the total error based on the
individual errors.  Since the individually measured errors are not
separable, the total error cannot be calculated directly; this
value is an approximation based on the measured values.}
\end{deluxetable}

The intrinsic errors are nonzero due to the nature of the abundance
determination.  The program must first determine a cross-correlation
velocity and continuum level in order to shift the data to match the
model spectra.  The finite resolution of the data and model spectra,
as well as the finite grid of velocity shifts and input model
abundances, occasionally leads to small offsets in abundance
determinations.  The uncertainty introduced by these various fitting
steps, including the velocity and continuum shift, as well as other
sources of error, are then included in the observed intrinsic scatter.
Note, however, that this scatter is still quite small.

Since external values for T$_{eff}$ and log $g$ are adopted in
choosing a model, appropriate errors in these parameters
\citep[$\sigma_{\mathrm{T}_{eff}}$ = 200 K, $\sigma_{\mathrm{log}
g}$ = 0.4;][]{SSPP} need to be translated into errors on the abundances.
Since [Fe/H] and [C/Fe] are fit internally, using the external values
only as a starting point, the errors on these parameters do not need
to be considered.

To translate these errors, the same synthetic spectra used to
determine the intrinsic error were refit with incorrect stellar
parameters ($\pm$100, $\pm$200 K in T$_{eff}$, and $\pm$0.2, $\pm$0.4
in log $g$, which correspond to the line index errors and are related
to the spacing of the model grid).  The resulting combined errors for
[Fe/H] and [C/Fe] based on the uncertainties in these parameters
\textit{and} the intrinsic errors are shown in Table \ref{tbl:error}.
Total errors are then estimated from these individual sources by
assuming approximate Gaussian distributions.  Since the errors are
essentially constant across log $g$, the total errors are most
reasonably read from the bins of varying temperature, and are thus
less than $\sim$ 0.20 dex for [Fe/H] and $\sim$ 0.15 dex for [C/Fe],
for stars warmer than 4000 K.

\begin{figure}
  \figurenum{2}
  \plotone{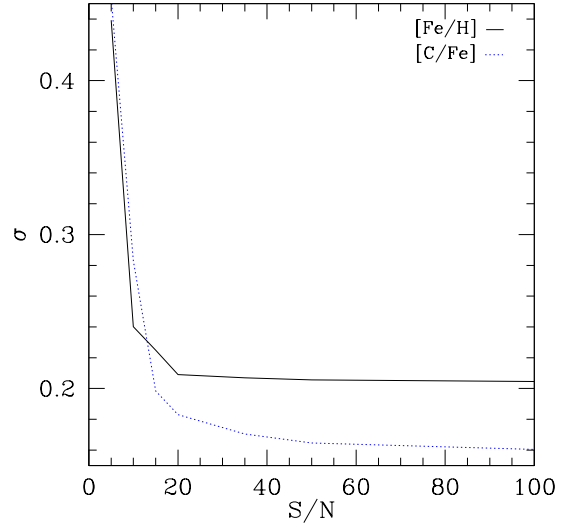}
  \caption{Total error for [Fe/H] (solid line) and [C/Fe] (dotted line) vs. signal-to-noise ratio.
    \label{fig:error}}
\end{figure}

The error in the abundances due to the changing signal to noise of the
data also needs to be considered.  To determine this, varying amounts
of noise were injected into the noise-free synthetic spectra, which
were run through autoMOOG again.  This error combined with the
previous errors (assuming $\sigma_{\mathrm{[Fe/H]}} = 0.20$ dex and
$\sigma_{\mathrm{[C/Fe]}} = 0.15$ dex) are shown in Figure
\ref{fig:error}.  The total errors are then less than 0.25 dex for S/N
$\gtrsim$ 10 for both [Fe/H] and [C/Fe].

\section{VALIDATION}
Validation of the autoMOOG approach requires comparison of the
results, as obtained from medium-resolution spectra, with stars that
have abundances determined from high-resolution analyses.  Differences
between high-resolution and lower-resolution analyses, such as
differences in adopted stellar parameters, different lines used for
individual abundances, and differences in fitting techniques, may all
lead to small scatter between results.  However, lower-resolution
techniques are often empirically calibrated to higher-resolution
studies, which somewhat mitigates this scatter.  Comparison of the autoMOOG
technique with other methods designed for lower-resolution spectra,
such as the line index method, would also be valuable, particularly
since they would be looking at the same features.  These comparisons
are performed for two sets of data, as described below.

\subsection{The Data}
The first set of stars analyzed are from the catalog of
\citet[hereafter BM]{BandM}, originally identified as metal-weak
candidates in the southern sky.  This sample ranges from around [Fe/H]
= $-2.5$ up to solar metallicity, and thus represents a modestly
metal-poor, and predominantly carbon-normal, sample.  The
high-resolution values used here come from the literature compilation
of \cite{Cayrel}.  The medium-resolution ($\sim1$ \AA, S/N $\sim 100$)
spectra that are being fit are described in T. C. Beers et al (2009), in
preparation.

\begin{figure}
  \figurenum{3}
  \plotone{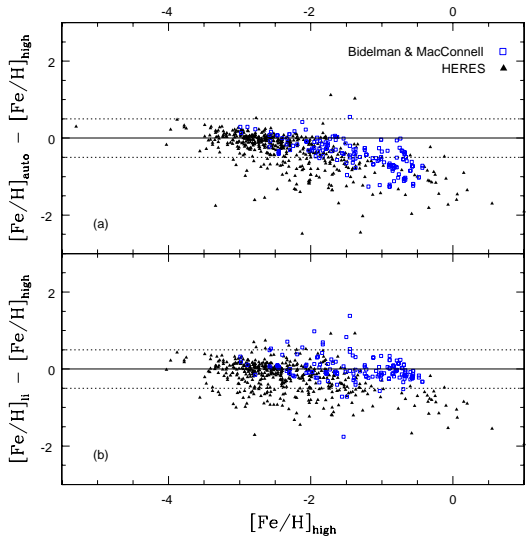}
  \caption{Comparisons of the metallicity determined from (a) autoMOOG
  and (b) the line index method to the high-resolution results.  In
  both cases, the higher metallicity, carbon-normal BM stars are
  plotted as open squares, while the more metal-poor, occasionally
  carbon-rich HERES stars are plotted as filled triangles.  The solid
  line corresponds to agreement with the various abundances.  The
  dashed lines to either side correspond to a difference of $\pm$0.5
  dex.
    \label{fig:fehcomp}}
\end{figure}

The other data are a sample of metal-poor stars observed as
part of the Hamburg/ESO R-process Enhanced Star survey
\citep[HERES;][]{HERES}.  Initially selected from the full Hamburg/ESO
and HK surveys, the HERES project obtained moderate-resolution
($\sim2$ \AA) follow-up spectroscopy for metal-poor giant candidates.
For bright stars confirmed to have [Fe/H] $< -2.5$,
higher-resolution ($R=20,000$, S/N $\sim 50$, $\lambda = 3760 - 4980$
\AA) spectroscopy was obtained \citep{Barklem, Luc}.  Their goal was
to identify the small fraction of these stars that showed large enhancements of
r-process elements, so that investigations into the nature of this
process could be conducted.  As a consequence of these observations,
several hundred metal-poor stars, many of them carbon-enhanced, were
observed at moderate and high resolution.

These then make ideal samples for the present analysis, as the
medium-resolution data can be run through autoMOOG, and the
high-resolution results are available for the purposes of comparison.

\subsection{Results}
Due to the high quality spectra available for the majority of the
medium-resolution data, metallicities and carbon abundances were
determined for nearly all the stars in the samples.  A total of 913
stars were fit, including 529 from the BM sample and an additional 384
from the HERES sample.  In some cases, stars had no metallicity
determination available from the line index method.  In addition, many
of the BM stars lacked high-resolution metallicities, and almost none
had high-resolution carbon abundances.  In all of these cases,
abundances were obtained using autoMOOG, but obviously no comparison
was possible.

When comparing autoMOOG results for metallicity with the
high-resolution analysis (Figure \ref{fig:fehcomp}a), it can be seen
that there exists a generally good agreement at low metallicities,
with a small scatter and no appreciable offset.  Due to issues of
continuum placement in the normalization of higher metallicity ([Fe/H]
$\gtrsim -2.0$) input spectra, an increasing underestimate of
metallicity is seen.

During the process of continuum subtraction and spectral normalization
prior to the running of autoMOOG, a continuum is placed along the
apparent continuum of a stellar spectrum.  For high-metallicity stars
observed at low spectral resolution, the multitude of metallic lines
causes a suppression in the apparent continuum, masking the true
continuum level.  Continuum subtraction techniques will
improperly normalize such spectra, and subsequent lines fit within
autoMOOG, such as calcium, will then appear weaker than they truly
are, resulting in a general underestimate in the metallicity
estimates.

\begin{figure}
  \figurenum{4}
  \plotone{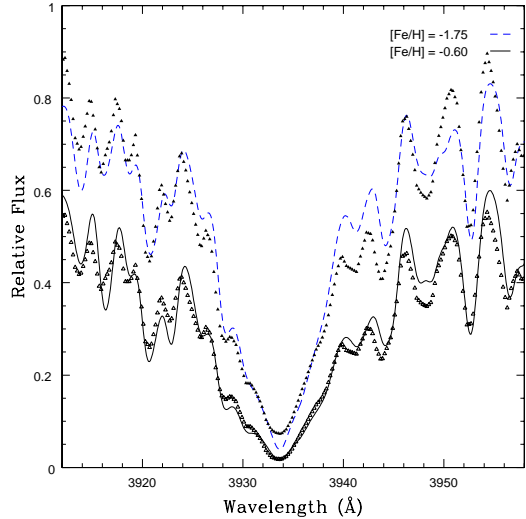}
  \caption{Synthetic spectrum near the Ca \textsc{II} K line assuming [Fe/H] =
  $-1.75$ (dashed line) and [Fe/H] = $-0.60$ (solid line) for a star of
  T$_{eff}$ = 4800, log $g$ = 2.2.  Overplotted are points
  corresponding to data from a star in the BM sample, using the data
  created after the continuum fit (filled triangles) and that same
  data scaled to match the solid line (open triangles).  Note the
  different apparent continuum levels for the metal-poor spectrum
  vs. the metal-rich spectrum.
    \label{fig:cakcomp}}
\end{figure}

This effect is demonstrated in Figure \ref{fig:cakcomp}.  Synthetic
spectra are plotted for models with T$_{eff}$ = 4800 K and log $g$ =
2.2, with possible metallicities of [Fe/H] = $-1.75$ (dashed line) and
[Fe/H] = $-0.60$ (solid line).  Overplotted on these synthetic spectra
are the data from star CD-57:949, taken from the BM sample.  The
filled triangles correspond to the flattened, normalized spectrum as
produced from the continuum fit to the data.  The open triangles, on
the other hand, are the same data after application of a scaling
factor selected to best match the higher metallicity synthetic
spectra.  Based on the input, filled triangle spectra, autoMOOG
appropriately selected a metallicity of [Fe/H] = $-1.75$ for this
star.  The fit is far from perfect, but is the best fit to the input
spectrum.  The filled triangles are clearly better matched by the
dashed line than they are by the solid line.

However, the higher metallicity synthetic spectrum, at [Fe/H] =
$-0.60$, is close to the actual metallicity of CD-57:949, as
determined from high-resolution analyses.  If the data were more
appropriately scaled, as with the open triangles, it would better
match this spectrum.  Even in this rough simulation, it is fairly
clear that the open triangles match the solid line much better than
the filled triangles match the dashed line.  However, from the data
supplied to autoMOOG, an incorrect, and always lower metallicity,
synthetic spectrum is chosen as the best fit. This problem becomes
more severe at higher metallicities as the apparent continuum is
further suppressed relative to the true continuum, resulting in larger
deviations to lower selected metallicities.  Thus, this is a problem
with the input spectra, not with autoMOOG itself.  With properly
continuum-subtracted spectra, autoMOOG will likely produce results
with similar scatter seen at low metallicity.

Efforts have been made to include renormalization within autoMOOG to
attempt to deal with this problem.  While renormalization does show
some promise, its results are currently erratic.  Future versions of
autoMOOG will likely include renormalization, but it will take
considerable modifications to the code before this becomes a stable
feature.

The line index results (Figure \ref{fig:fehcomp}b) exhibit more
scatter than autoMOOG at low metallicity, but similar behavior at high
metallicity.  For the line index approach at high metallicity, this is
caused in part by the same continuum issues mentioned above but also
by saturation of the KP index.  This can occasionally be minimized by
substitution of other metallicity determination methods designed
specifically to handle higher metallicity stars, such as the
auto-correlation function suggested by \cite{Rat} and explored by
\cite{B99}.  This accounts for the smaller deviations from the
high-resolution results of the BM stars, but not the HERES stars, seen
in the line index comparison plot.  Also note that the stars with
[Fe/H] $\lesssim -1.0$ that show the largest deviations in Figure
\ref{fig:fehcomp}a are simply not present in Figure
\ref{fig:fehcomp}b, as the line index method failed to determine any
metallicity for these stars.

As such, metallicities determined with autoMOOG are generally \emph{as
good as} those determined using existing techniques, and somewhat
better at the lowest metallicities ([Fe/H] $< -3.0$).  Since this
method was designed for use with VMP stars, it is then better than
existing techniques at accurately determining metallicities.  However,
due to the difficulties at higher metallicity, it should not be used
blindly with samples that may contain stars with metallicity greater
than about [Fe/H] = $-2.0$.

\begin{figure}
  \figurenum{5}
  \plotone{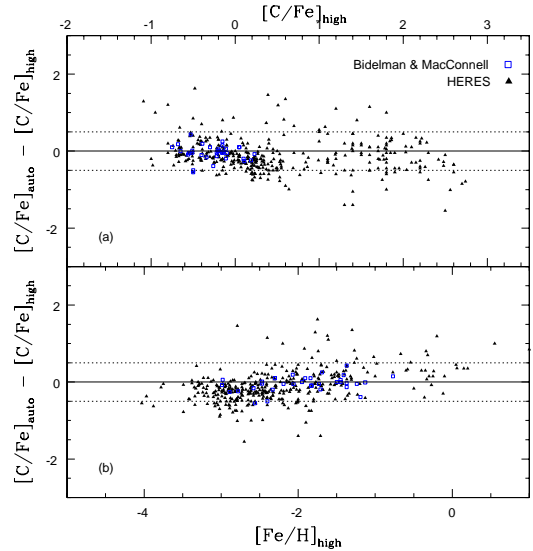}
  \caption{Comparisons of the carbon abundance, [C/Fe], determined
  from autoMOOG to that determined from high-resolution analyses,
  plotted vs. the high-resolution (a) [C/Fe] and (b) [Fe/H].  The
  points are the same as in Figure \ref{fig:fehcomp}.  Upper limits
  are not plotted, only determined values.
    \label{fig:cfecomp}}
\end{figure}

Much better agreement is seen when comparing the autoMOOG fits on
carbon, [C/Fe], to the high-resolution abundances (Figure
\ref{fig:cfecomp}).  The only stars plotted are those for which fits
(as opposed to upper limits) have been obtained and for which
high-resolution abundances are available.  When plotted as a function of
[C/Fe] (Figure \ref{fig:cfecomp}a), it is seen that there is very
little scatter at lower [C/Fe] and only slightly larger scatter as the
carbon abundance is increased.  At higher carbon abundances, the added
scatter is likely due to the presence of other strong carbon features
in the vicinity of the $G$ band, or the decreased sensitivity of the $G$
band as it begins to saturate.

Examination of the carbon fits as a function of metallicity (Figure
\ref{fig:cfecomp}b) indicates little scatter, but a small slope seems
to be present.  This slope is not real, but is caused by a jump at
[Fe/H] $\sim-2.5$, with a slight underestimation of [C/Fe] at lower
metallicities, and a slight overestimation at higher metallicities.
The underestimation at low metallicity is likely caused by the use of
different carbon features at high and low resolution, which are
affected differently by model atmosphere parameters (i.e.,
one-dimensional versus three-dimensional and LTE versus NLTE).  This
then should extend to higher metallicities as well, for a global
underestimation of around 0.25 dex.

The slight overestimation of [C/Fe] at high metallicity is likely
caused by the same difficulties that have already been seen at high
metallicity, although the effect is much weaker here.  The reason for
this weaker effect is that the carbon fits are actually a fit of
[C/M], where M is the scaled-solar model abundance.  To ultimately
derive [C/Fe], this model-dependent carbon value needs to be offset by
the deviation of iron from the model.  This offset is at most half of
the distance between models in our grid (i.e., $\Delta$[Fe/H] $\leq
0.25$ dex), and thus should not significantly affect the final results,
but causes this slight relative overestimation.

In addition, the temperatures used were determined from photometric
colors.  At times, the parameters adopted for the high-resolution
analysis vary from these values, occasionally up to a difference of
nearly 1000 K, although generally only by a few hundred kelvin.  As
seen in Section 3, such offsets in temperature could lead to offsets in
metallicity of $\sim$0.15 dex and $\sim$0.10 dex in [C/Fe].  Due to the
inconsistent nature of these offsets, this could also lead to enhanced
scatter in any of the comparisons to high resolution.

It is also possible to use these results to perform an internal
consistency check on the method.  Since a number of stars from the BM
sample have been observed multiple times at medium resolution, the
abundances determined for each individual spectra can then be compared
to see if autoMOOG determines the same relative abundances for a given
star, regardless of the spectra used.  For each star, an average
metallicity and carbon abundance are calculated from the available
results, and then the residuals between each individual abundance and
the average are calculated.

For both metallicity and carbon abundance, a very small scatter around
zero of $\sim$ 0.02 dex is observed, well within the intrinsic errors,
showing the overall stability of autoMOOG.  The scatter for [Fe/H] at
high metallicities may be slightly larger than elsewhere.  This is
assumed to be a result of the exact continuum fit used for each
spectra, and is thus associated with the general high-metallicity
difficulties.  Even so, this scatter is still quite small, increasing
by no more than 0.01 dex.

\section{UPPER LIMITS}
\begin{figure}
  \figurenum{6}
  \plotone{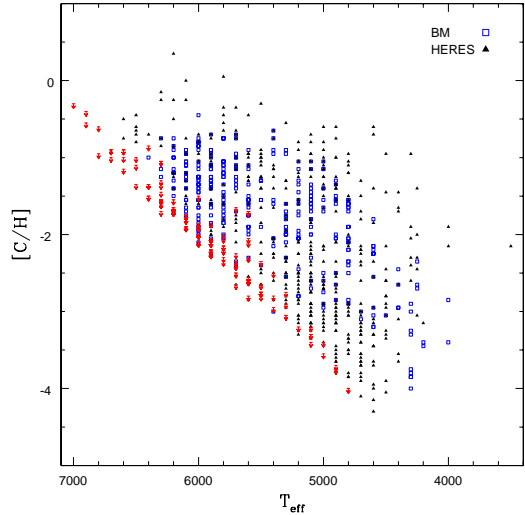}
  \caption{Carbon abundances ([C/H]) plotted vs. effective
  temperature.  Plotted are the abundances for the BM sample (open
  squares) and the HERES sample (filled triangles).  All upper limits
  are denoted with an arrow.
    \label{fig:cvt}}
\end{figure}

For the purpose of this work, a $G$ band is considered detectable if its
deflection from the continuum is at least 5\% the value of the
continuum at that position.  Upper limits for [C/Fe] were determined
by varying abundances of synthetic spectra and comparing their
strengths against an essentially ``zero'' carbon abundance spectrum
which was used to set the continuum level.  A surface in T$_{eff}$-log
$g$ space was then fit to these values.  Stars with carbon abundances
below this surface were reassigned with the value of the upper limit
at that position, and flagged appropriately.  The approximate position
of this cutoff can be seen in Figure \ref{fig:cvt}.

\begin{figure}
  \figurenum{7}
  \plotone{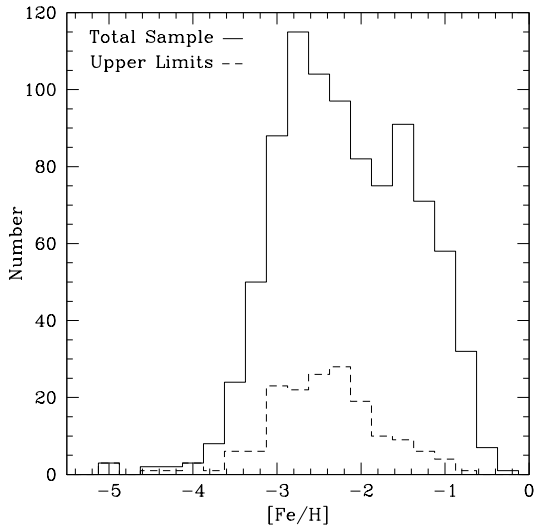}
  \caption{Metallicity distribution function for the full HERES and BM
  sample (solid line) and those stars for which only an upper limit on
  carbon have been determined (dashed line).
    \label{fig:fhistup}}
\end{figure}

When the $G$ band can be properly fit, appropriate [C/Fe] values are
generally determined.  It is, however, useful to explore under what
physical conditions only an upper limit is obtained.  The
detectability of the $G$ band relative to temperature and metallicity is
shown in Figures \ref{fig:cvt} and \ref{fig:fhistup}, respectively.

In Figure \ref{fig:cvt}, the carbon abundance, [C/H], is plotted
versus effective temperature using the values from autoMOOG.  The
stars in these samples cover a fairly wide range of temperatures, from
4000 K to 7000 K.  Both samples of stars cover about the same
temperature range.

The $G$ band is a molecular feature, which, for the same abundance of
carbon, will appear stronger at lower temperatures.  As temperature is
increased, the $G$ band weakens until it is no longer detectable.  The
point where this happens is obviously dependent on the level of carbon
enhancement, [C/H], and is easily seen in the plot.  [C/H] is used for
this analysis as it is solely a measure of the carbon abundance,
without any implicit assumptions about overall metallicity, which,
along with the temperature, is primarily what the strength of the $G$
band depends on\footnote{The $G$ band also depends weakly on surface
gravity.  In addition, nitrogen and oxygen may lock up some of the
carbon in CN or CO molecules, which cannot then form CH.  However,
this is likely to produce only a negligible change in the strength of
the $G$ band at this precision even in cool O- or N-rich stars, where
such effects would generally be strongest.}.  From the line of
detectability, it can be seen that for cool stars, the $G$ band can
essentially always be detected.  However, for warm stars, especially
above 6000 K, carbon abundances must be fairly high to be detected.
Indeed, for temperatures above 7000 K, it would be difficult to
measure a carbon abundance at all, no matter how high its true
abundance.  In part, this is due to the apparent upper limit on [C/H].

Although there are a handful of stars with slightly higher carbon
abundances, a general limit to the allowed carbon abundances falling
at [C/H] $\sim -0.5$ can be seen, which is higher than some have found
\citep[$-1.0$ and $-0.7$ :][respectively]{Ryan,Frebel}, while lower
than others \citep[0.0 :][]{Lucatello,Aoki}.  It is likely that
the true location of this upper limit is temperature dependent, with
cooler stars exhibiting a lower value, and warmer stars a higher
value.  This may be visible in Figure \ref{fig:cvt}, and can further
be supported by the previously mentioned results, and is presumably
due to evolutionary effects.

While the two samples we have examined do not fully complement each
other, as seen in their combined Metallicity Distribution Function
(MDF; Figure \ref{fig:fhistup}), their results can still be used to
analyze the detectability of the G band with respect to metallicity.
Figure \ref{fig:fhistup} shows the MDF of the full samples of
stars, and the MDF of stars with only a measured upper limit.

This second set of stars overall traces the shape of the full MDF,
although accounting for slightly larger percentages of stars at lower
metallicities.  This is not surprising, as metal-poor stars have, by
definition, low abundances and weak lines.  However, due to the high
quality of these samples, carbon abundances are able to be determined
for a significant fraction of stars, even at the lowest metallicities.

\section{CONCLUSIONS AND FUTURE WORK}
We have developed a new method for the determination of [Fe/H] and
[C/Fe] for stars using automated spectral synthesis.  This method
appears to not be affected by the presence of additional C-related
features in the spectra of some stars that has been claimed by
\citet{Cohen} to limit the accuracy of the line index method for CEMP
stars, and does not require the massive calibration required by that
method.  autoMOOG is significantly quicker and more consistent than
spectral synthesis by hand, which in general could vary from
individual to individual, and even for one individual over the course
of time.

We have explored the behavior of this method for two samples of stars
with both medium-resolution and high-resolution spectroscopy
available.  Using this method on two samples of stars which cover a
wide range of parameter space, we are able to see exactly how this
method behaves.  For stars with [Fe/H] $\le -2.0$, both the
metallicity, [Fe/H], and carbon abundance, [C/Fe], where detectable,
are able to be successfully recovered.  Despite the known problems
with their metallicities, the carbon abundances for the more
metal-rich stars are similarly recovered.  As this method was designed
and optimized primarily for use with metal-poor stars, it should not
be used indiscriminately with samples containing more metal-rich
samples, as these stars will contaminate metal-poor studies due to the
metallicity underestimation.  In future versions of this technique, we
will attempt to resolve the issues present at high metallicity through
implementation of various methodologies currently under development.
However, at least for stars with [Fe/H] $\le -2.0$, it is more
consistently accurate (with errors of around 0.2 dex for both [Fe/H]
and [C/Fe]) than previous abundance methods, especially at the
lowest metallicities, where these other methods break down.

Although we do not show it here, this program is capable of fitting
any element from any detectable line.  Thus, it would be possible to,
for instance, measure the abundance (or an upper limit) of s-process
elements such as barium (4554 \AA) or strontium (4078 \AA), expected
to be enhanced for many CEMP stars.  In this way, we could roughly
separate CEMP stars into the CEMP-s and CEMP-no classes \citep{BaC}.
However, detailed studies of these fits have not yet been performed.
In particular, no similar upper limit mechanism is in place for these
other lines.  For the moment, fitting other features should only be
done for very rough estimates, and should not be used for any detailed
science, but can be used for identification of likely neutron-capture
enhanced targets for high-resolution follow-up.  Future versions of
autoMOOG will allow fitting for a larger variety of lines with more
robust detection and saturation algorithms, as well as internal
fitting of effective temperature and surface gravity, to provide less
dependence on external techniques.

The authors are grateful to John Norris for permitting access to his
spectroscopic observations of the Bidelman \& MacConnell candidates
prior to publication.  B.M., T.C.B., and T.S. acknowledge partial
support from grants AST 00-98508, AST 00-98549, AST 04-06784, and AST
07-07776, as well as PHY 02-16783 and PHY 08-22648, Physics Frontier
Centers/JINA: Joint Institute for Nuclear Astrophysics, awarded by the
US National Science Foundation.  S.R. and V.P. acknowledge partial
financial support from the Brazilian institutions FAPESP, CNPq and
Capes.  S.L. acknowledges partial support from INAF cofin 2006 and the
DFG cluster of excellence 'Origin and Structure of the Universe'
(www.universe-cluster.de)

\end{document}